\documentclass[twoside,onecolumn,9pt]{article}
\usepackage{extsizes}
\usepackage[super,sort&compress,comma]{natbib} 
\usepackage[version=3]{mhchem}
\usepackage[left=1.5cm, right=1.5cm, top=1.785cm, bottom=2.0cm]{geometry}
\usepackage{balance}
\usepackage{widetext}
\usepackage{times,mathptmx}
\usepackage{sectsty}
\usepackage{graphicx} 
\usepackage{lastpage}
\usepackage[format=plain,justification=raggedright,singlelinecheck=false,font={stretch=1.125,small,sf},labelfont=bf,labelsep=space]{caption}
\usepackage{float}
\usepackage{fancyhdr}
\usepackage{fnpos}
\usepackage[english]{babel}
\usepackage{lipsum}
\usepackage{array}
\usepackage{droidsans}
\usepackage{charter}
\usepackage[T1]{fontenc}
\usepackage[usenames,dvipsnames]{xcolor}
\usepackage{setspace}
\usepackage[compact]{titlesec}
\usepackage{epstopdf}
\usepackage{amsmath}

\definecolor{cream}{RGB}{222,217,201}

\begin{document}

\noindent\LARGE{\textbf{Very large scale characterization of graphene mechanical devices using a colorimetry technique$^\dag$}} \\
\vspace{0.3cm} \\
\noindent\large{Santiago Jose Cartamil-Bueno,$^{\ast}$\textit{$^{a}$} Alba Centeno,\textit{$^{b}$} Amaia Zurutuza,\textit{$^{b}$} Peter Gerard Steeneken,\textit{$^{a}$} Herre Sjoerd Jan van der Zant,\textit{$^{a}$} Samer Houri $^{\ast}$\textit{$^{a}$}} \\

\section*{}
\vspace{-1cm}

\footnotetext{\textit{$^{a}$~Kavli Institute of Nanoscience, Delft University of Technology, Lorentzweg 1, 2628 CJ Delft, The Netherlands.}}
\footnotetext{\textit{$^{b}$~Graphenea SA, 20018 Donostia-San Sebasti\'an, Spain.}}
\footnotetext{\textit{$^{\ast}$~E-mail: s.j.cartamilbueno@tudelft.nl, s.houri@tudelft.nl}}

\footnotetext{\dag~Electronic Supplementary Information (ESI) available: Fabrication of SLG and DLG drumheads, derivation of the expression for adhesion energy, Raman spectroscopy of suspended and collapsed DLG drumheads, theoretical limit of large-diameter drums before collapse, stuck drums and the diameter-dependence of their occurrence probability, complete dataset of yield for SLG and DLG samples. See DOI: 10.1039/b000000x/}

\sffamily{\textbf{We use a scalable optical technique to characterize more than 21000 circular nanomechanical devices made out of suspended single- and double-layer graphene on cavities with different diameters ($D$) and depths ($g$). To maximize the contrast between suspended and broken membranes we used a model for selecting the optimal color filter. The method enables parallel and automatized image processing for yield statistics. We find the survival probability to be correlated to a structural mechanics scaling parameter given by $D^4/g^3$. Moreover, we extract a median adhesion energy of $\Gamma =$ 0.9~J/m$^2$ between the membrane and the native SiO$_2$ at the bottom of the cavities.}}\\

\rmfamily 


Graphene, the 2D allotrope of carbon and a monolayer of graphite, is expected to reach the markets \cite{Alcalde2013,Ferrari2014,Zurutuza2014} after years from its isolation in 2004 \cite{Novoselov2004}. Of all 2D materials studied at present, graphene is the strongest with a fracture strength of 130~GPa and a stiffness of 1~TPa in its pristine form after exfoliation\cite{Lee2008a}. Together with its excellent electronic properties \cite{Bolotin2008b,Du2008} freestanding graphene promises to be the ultimate material for micro-/nano-electromechanical systems (M/NEMS), showing potential as RF oscillators\cite{Chen2013, Houri2017}, pressure sensors \cite{Smith2013,Dolleman2015}, and electromechanical switches \cite{Liu2014}.

A promising means for industrial scale production of graphene and other 2D materials is the use of the chemical vapour deposition (CVD) technique on catalytic foils such as copper \cite{Li2009,Li2016} or germanium \cite{Lee2014}. However, the fabrication of suspended membranes requires a complicated transfer process that impacts device performance negatively \cite{Zande2010} and results in a low yield \cite{Suk2011}.
Studies performed so far on CVD graphene have shown large variations in its physical properties such as Young's modulus \cite{Lee2013}, mass density or stress \cite{Zande2010,Barton2011} and permeability \cite{Celebi2014}. To get a better insight into these variations, the mechanical properties of CVD graphene should be determined not only at device level, but also on a very-large-scale integration (VLSI) level for commercial applications.

Although inspection via optical microscopy could be used for such a massive characterization, the monoatomic thickness of graphene leads to a low optical reflectivity that makes it hard to study. Visibility under an optical microscope can be enhanced by selecting the thickness and optical properties of the underlying substrate \cite{Blake2007}.
However, this approach has not yet been applied in the case for suspended graphene.
Statistical studies were performed with laser interferometry \cite{Zande2010,Barton2011}, Raman spectroscopy \cite{Suk2011,Metten2014,Shin2016,Wagner2017}, atomic force microscopy \cite{Lee2013,Hwangbo2014}, and electrical modulation \cite{Arjmandi-Tash2017}, but such approaches tend to be complicated and time consuming hence preventing a successful commercialization of graphene mechanical devices.

In this work, we use a scalable colorimetry technique \cite{Cartamil-Bueno2016} to measure yield statistics on more than 21000 graphene drumheads of different diameters and cavity depths (several substrates) for both CVD single-layer (SLG) and double-layer graphene (DLG). We also use an optical model to optimize the reflectance contrast between suspended and broken devices. With these tools, we obtain the survival probability of SLG and DLG samples, and find a scaling rule that governs the yield of suspended CVD graphene membranes.

Device fabrication starts by etching circular holes into the thermally-grown SiO$_2$ layer on a silicon chip. A total number of 3016 circular cavities are produced on each sample substrate, having diameters ranging from 2 to 20~$\mu$m. Several chips are fabricated with cavity depths of 285, 570, 630 and 1140~nm, and two additional samples of 630~nm are coated with a thin layer of Au/Ti (100~nm/10~nm).
Afterwards, the samples are sent to Graphenea to transfer the CVD SLG and CVD DLG. For more information about the fabrication, see Electronic Supplementary Information.

\begin{figure}[H]
	\centering
	\includegraphics[]{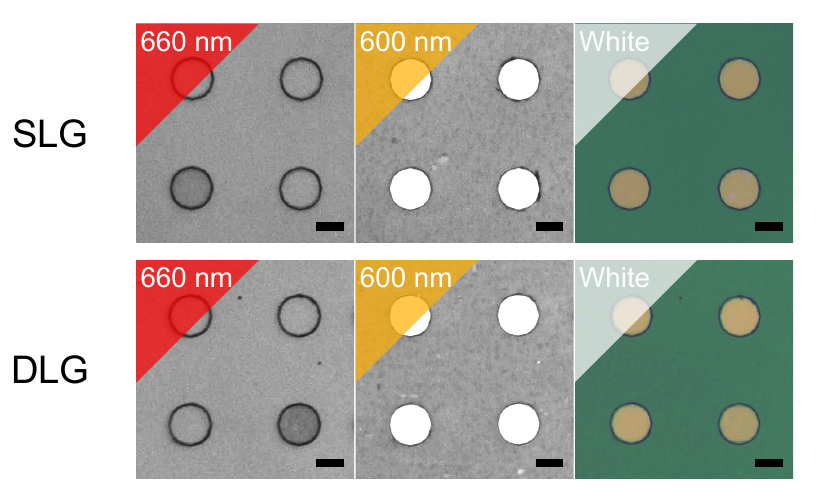}
	\caption{Visualization of suspended graphene drumheads using color filters.
		Optical microscope images with narrow band illumination (660~nm and 600~nm) and white light for CVD single layer (SLG) and double layer (DLG) graphene suspended on Si/SiO$_2$ cavities 1140~nm deep. The suspended 15~$\mu$m-in-diameter graphene membranes appear darker than the broken ones when using a color filter of 660 nm. Scale bars are 10~$\mu$m.}
	\label{fgr:fig1}
\end{figure}

Upon inspection of the suspended structures with an optical microscope, the surviving devices prove to be difficult to identify as they offer very low optical contrast with broken drums. This low contrast is seen in the rightmost panels of Figure~\ref{fgr:fig1} for cavities of 1140~nm in depth in two different samples (SLG top, and DLG bottom), in which under white light illumination a suspended drumhead is indiscernible from the broken ones.
This is due to interference of the light being reflected from the Si substrate and that being reflected from the graphene layer \cite{Kim2015}. This interference can be either constructive or destructive depending on the wavelength $\lambda$ and effective cavity depth $g$ (distance between the membrane and the bottom of the cavity). In the case of white light illumination, these constructive and destructive interference contributions average out, thus reducing the observed contrast such that suspended drums become indiscernible.

A judicious selection of the illumination wavelengths can increase the contrast and therefore the visibility of the suspended drums. A similar approach was taken to make supported graphene on SiO$_2$ more visible \cite{Blake2007}. This is also shown in the leftmost panels of Figure~\ref{fgr:fig1} where using a 660~nm color filter (10~nm of bandwidth) enhances the contrast between suspended and broken drumheads, thus enabling their identification.
On the other hand, this enhanced contrast is once again lost if a filter wavelength of 600~nm is chosen. As seen in the figure, the behavior is common to both SLG and DLG, although the DLG membrane absorbs more light thus giving a higher contrast between suspended and broken devices.

For a given cavity depth, the relation between reflected contrast and light wavelength can be understood with a Fresnel-law model with three optical layers (air, graphene, air) interfacing a reflective substrate (silicon). The reflectance of the suspended membrane can be written as: 

\begin{equation}
\begin{split}
R = &  | 2r_1e^{i\phi_2}\sin \left(\phi_1 \right)  + r_2e^{-(\phi_1+\phi_2)} \\
	&- \frac{r_1^2r_2e^{i(\phi_1-\phi_2)}}{e^{i(\phi_1+\phi_2)} + r_1^2e^{-i(\phi_1-\phi_2)} 
	+ 2r_1r_2e^{-i\phi_2}\sin \left(\phi_1 \right) } |,
\label{eqn:Eq1}
\end{split}
\end{equation}

\noindent
where $r_1$ and $r_2$ are the Fresnel reflection coefficients of air-graphene and air-silicon interfaces, respectively. $\phi_1$ and $\phi_2$ are the phase changes induced by the optical path through the graphene and the cavity respectively, $t_G$ is the graphene layer thickness, and $\lambda$ is the optical wavelength \cite{Blake2007}.

To find the optimum wavelength to differentiate suspended from failed membranes, we use the Weber contrast between the reflectance of a suspended membrane and the silicon substrate (representative of a broken drum). The Weber contrast, $C$, is defined as:

\begin{equation}
C = \left| \frac{R-R_{Si}}{R_{Si}} \right|.
\label{eqn:Eq2}
\end{equation}

\begin{figure}[H]
	\centering
	\includegraphics[]{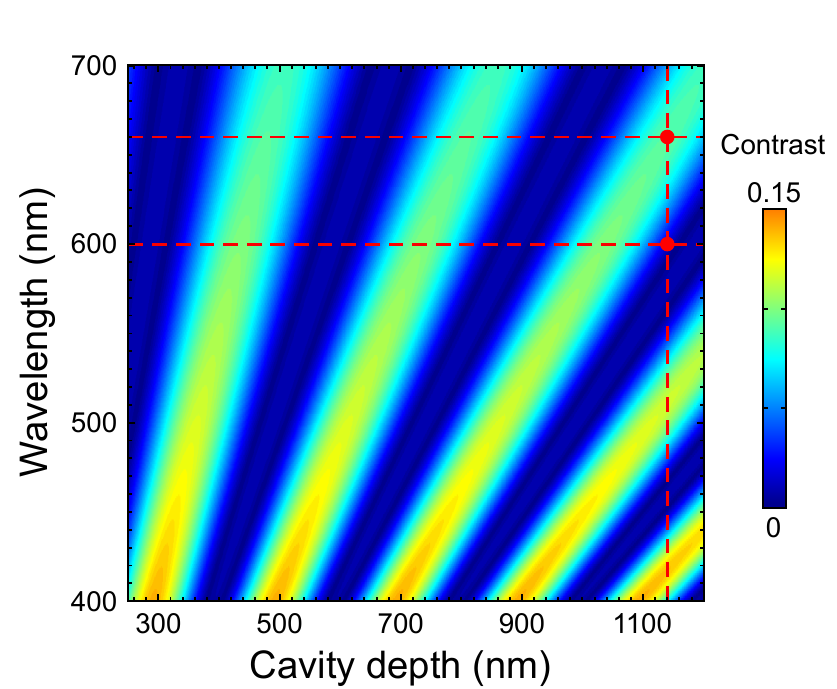}
	\caption{Calculated optical Weber contrast ($C$) as a function of the cavity depth and light wavelength for freestanding SLG.
		For a given cavity depth, the contrast between a suspended membrane and the silicon substrate (broken device) can be maximized by choosing the right color filter. The intersections of the dashed red lines indicate the expected contrasts for the drumheads shown in Figure~\ref{fgr:fig1} (depth of 1140~nm). The refractive index of graphene is taken to be $n_G = $2.6-1.3i.}
	\label{fgr:fig2}
\end{figure}

\noindent
Figure~\ref{fgr:fig2} plots the calculated values of $C$ for different cavity depths and wavelengths in the case of SLG. The reflected contrast value can be maximized for a given cavity depth by selecting the optimal illumination wavelength. For instance, the freestanding membranes of the samples in Figure~\ref{fgr:fig1} would be more visible when using color filters of 420~nm, 510~nm or 660~nm (dotted vertical red line at cavity depth = 1140 nm). On the contrary, illumination of 600~nm would lead to a lower contrast, as seen experimentally.

\begin{figure}[H]
	\centering
	\includegraphics[]{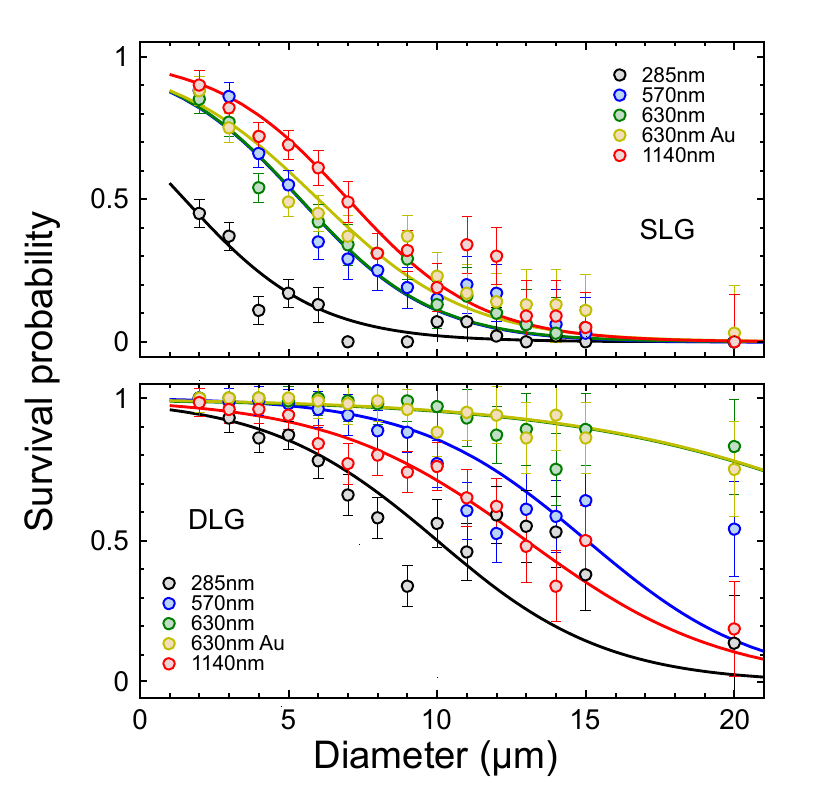}
	\caption{Survival probability of suspended CVD SLG (top) and DLG (bottom) as a function of cavity depth and diameter. Sigmoid curves (solid lines) are added as a guide to the eye to underline the tendency of large drumheads and shallow cavities to fail. The survival probability is significantly higher for DLG samples. The legend values indicate the cavity depth $g$.}
	\label{fgr:fig3}
\end{figure}

Next, we perform a large-scale characterization of device yield by using the colorimetry technique. 
The measurements are done on samples that have undergone more than 2000 stress cycles consisting of changing the environment pressure from 1~bar to 10~mbar (pumping) and viceversa (venting).
For all cavity depths the optimum filter is selected using Figure~\ref{fgr:fig2}. Filtered microscopic images of large numbers of drums are made and analyzed. We calculate the yield as the ratio of surviving-to-total drums, we obtain the error bar as the standard deviation divided by the square root of the total number of devices, and we equate the yield fraction with survival probability by inserting the corresponding error bar. Figure~\ref{fgr:fig3} collects the yield for the SLG and DLG samples (top and bottom panel, respectively) for different cavity depths (in different colors). The solid lines are guide-to-the-eye sigmoid curves plotted to show the survival probability dependence of suspended membranes on the cavity diameter. Three trends are clearly observable: the yield increases with deeper cavities, the yield reduces with increasing diameter, and the yield of DLG is higher than that of SLG. The only notable exception are the DLG samples with 1140~nm cavity depth that will be discussed below.

To quantitatively understand the dependence of the yield on the cavity diameter and depth, we hypothesize that failure is caused by the physical contact between the drums and the Si bottom of the cavity. The contact would take place when the membranes deflect downwards under the effect of a pressure difference between the inside and outside of the cavity.

In the above failure mode, the drum's deflection needs to be equal to the cavity depth.
By treating the drums as circular elastic membranes, it is possible to relate the maximum deflection of the centre of the membrane $\delta$ to an external force as:

\begin{equation}
F = \Delta P A_T = k_1 \delta + k_3 \delta ^3,
\label{eqn:Eq3}
\end{equation}

\noindent
where $F$ is the force causing the membrane to deflect, originating from the pressure difference $\Delta P$ acting on the drum's total area $A_T$. $k_1$ and $k_3$ are the linear and cubic stiffness of the circular membrane \cite{Boddeti2013,Liu2013,Cartamil-Bueno2016}.
Since we assume membrane behavior, the linear component of the stiffness is dictated by the pretension $T_0$ in the structure, $k_1=4\pi T_0$. The nonlinear component originates from the stretching of the membrane as it deforms, with $k_3 = \frac{32 \pi E_{2D}}{3 D^2 (1-\nu)}$. Here, $E_{2D}$ is the 2D Young's modulus of the graphene membrane taken to be 350~N/m for the SLG \cite{Lee2008a} and 700~N/m for DLG, $\nu = 0.16$ is the Poisson's ratio of graphene, and $D$ is the membrane's diameter.
With typical values of $k_1$ around tens of mN/m, it is clear that the nonlinear term dominates for deflections on the order of 100~nm and larger. Then for $\delta = g$, the correspondence between the cubic term and the pressure difference deforming the membrane is given by:

\begin{equation}
\frac{E_{2D}}{\Delta P} \approx \frac{3 (1-\nu)}{128} \frac{D^4}{g^3}.
\label{eqn:Eq4}
\end{equation}

\begin{figure*}[H]
	\centering
	\includegraphics[]{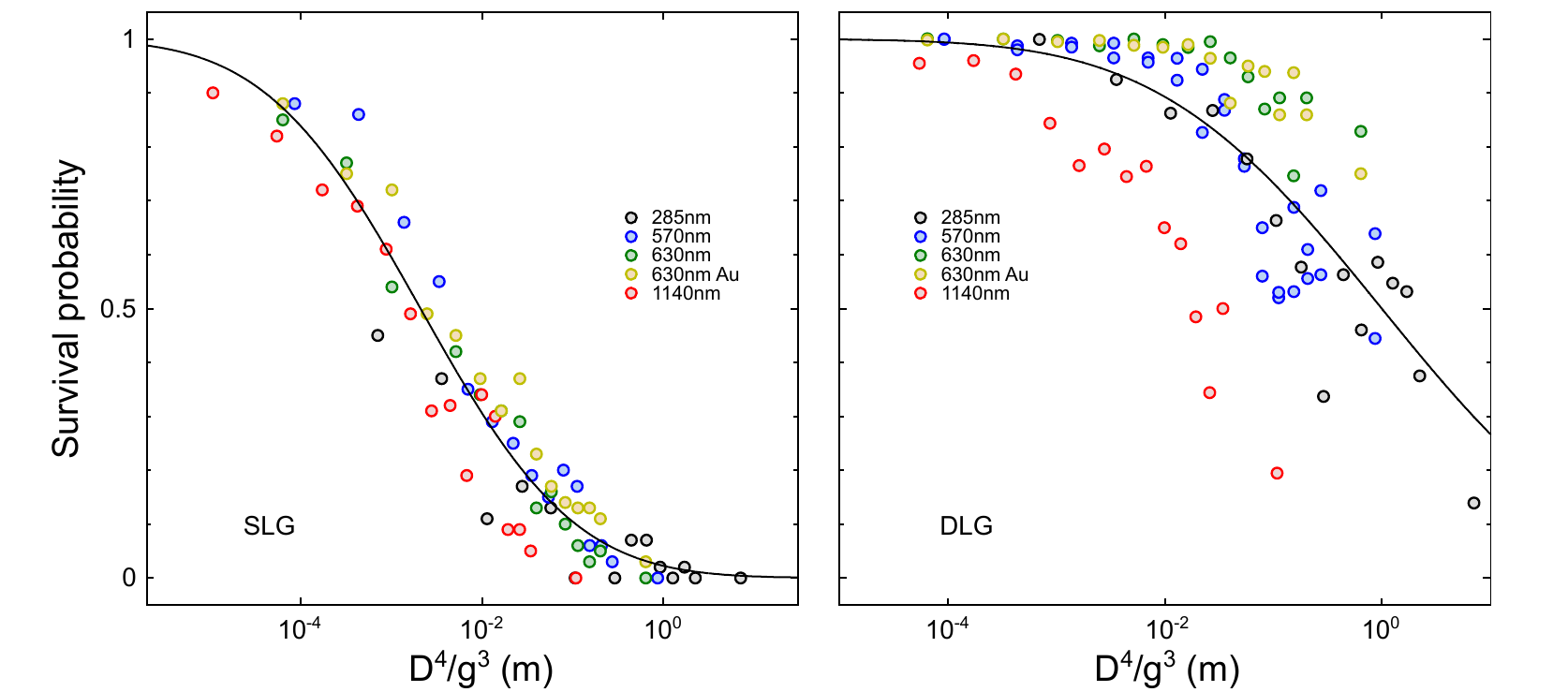}
	\caption{Fitted logarithmic-Gaussian distribution (solid lines) of the survival probability of 5715 suspended SLG drumheads (left) and 16079 suspended DLG drumheads (right) plotted against the cubic stiffness parameter $D^4/g^3$.}
	\label{fgr:fig4}
\end{figure*}

Therefore, for a certain pressure difference $\Delta P$ it is expected that the drums with $\frac{D^4}{g^3} < \frac{128E_{2D}}{3(1-\nu)\Delta P}$ will not touch the bottom of the cavity, and will thus survive. In Figure~\ref{fgr:fig4} we plot the yield of all the data for SLG (left panel) and DLG (right panel) as a function of the scaling parameter $D^4/g^3$. It is clear from the figure that the different samples tend to collapse on a single curve with a trend that can be fitted using a logarithmic Gaussian distribution (solid lines). From the fit we obtain using Equation~\ref{eqn:Eq4} the median value of $E_{2D}/\Delta P$ to be $4.1 \cdot 10^{-5}$~m for SLG and $2.2 \cdot 10^{-2}$~m for DLG. Thus, doubling the number of graphene layers shifts the median value of the scaling parameter by three orders of magnitude. The survival probability in Figure~\ref{fgr:fig4} follows a distribution that extends over 5 decades, whereas typical variations in the Young's modulus of CVD graphene are on the order of 10\% \cite{Lee2013}, and cannot therefore explain the span of the observed distribution.
However, since the membranes have different permeabilities, they will not experience the same $\Delta P$ during pressure cycling, which can explain the extracted ratio $E_{2D}/\Delta P$. Leakage could also account for some of the large 20~$\mu$m DLG membranes surviving the pressure test despite the predicted failure according to theory (Electronic Supplementary Information).

Moreover, the DLG sample with 1140~nm of cavity depth shows a larger failure probability for all diameters against the general trend. This fact could be an indication of a failure mechanism different from the collapse of the membrane and related to the cavity volume. Wherein, the volume plays an important role in the deflection of these devices due to the asymmetry between inflation and deflation of the membranes \cite{Cartamil-Bueno2016}.

\begin{figure}[H]
	\centering
	\includegraphics[]{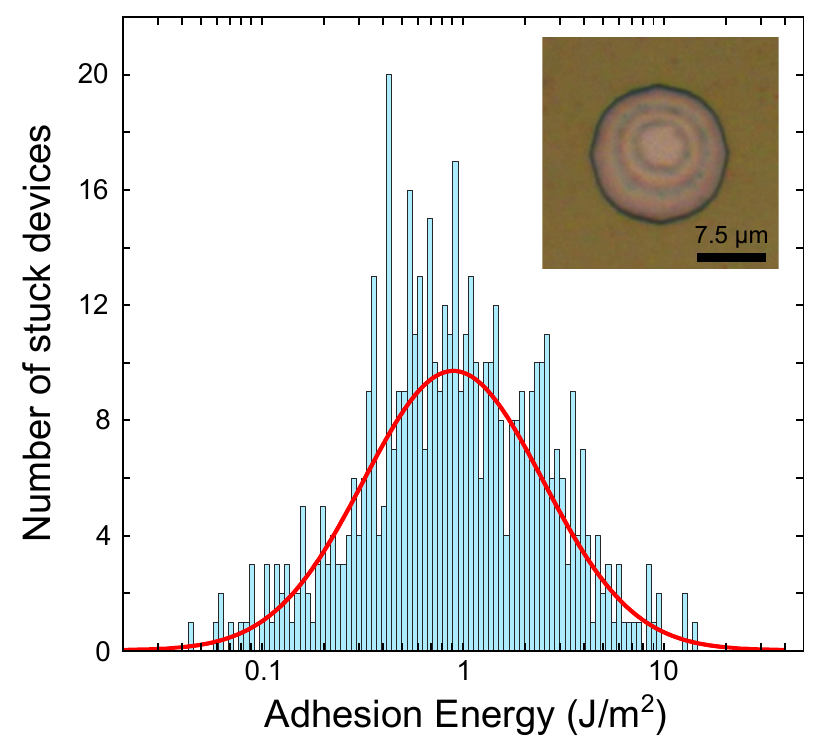}
	\caption{Histogram of the adhesion energy of 506 stuck DLG drumheads (all samples on SiO$_2$). The fitted Gaussian distribution appears as solid line. Inset: Optical image of a stuck drumhead. Scale bar is 15~$\mu$m.}
	\label{fgr:fig5}
\end{figure}

Besides broken structures, another failure mode was observed in a small subset of the membranes (5-30\% of the total devices per sample). These devices show fringes in the optical images as illustrated in the inset of Figure~\ref{fgr:fig5}, which indicate that the devices are permanently adhered (stuck) to the cavity bottom. The collapsed drum shown in the inset was also mapped using an atomic force microscope (AFM) as shown in the Electronic Supplementary Information.
Those stuck devices give an opportunity to explore graphene adhesion to surfaces. Curiously, although we observe such stuck devices for the DLG samples we observe no such features in the SLG samples, indicating that adhesion ruptures and destroys suspended SLG membranes upon contact with the substrate.

We can obtain information about the adhesion energy by analyzing the adhered area $A_S$. By assuming a perfect clamping and approximating the stuck area by a circle, it is possible to extract the adhesion energy through the equilibrium between the adhesion force and the restoring tension (Electronic Supplementary Information). This is given as:

\begin{equation}
\Gamma = 16E_{2D} \left( \frac{g}{D}\right)^4 \frac{2+\sqrt{\eta}}{\sqrt{\eta} \left(1-\sqrt{\eta}\right)^4}.
\label{eqn:Eq5}
\end{equation}

\noindent
where $\Gamma$ is the adhesion energy, and $\eta = \frac{A_S}{A_T}$ is the stuck-to-cavity area ratio.

The histogram of the extracted adhesion energy is shown in Figure~\ref{fgr:fig5} for a total of 506 DLG devices stuck to the bottom of their cavities. By fitting a logarithmic Gaussian distribution to the histogram (red line), we obtain a median value for the adhesion energy of 0.9~J/m$^2$.

The obtained value of adhesion energy for a pristine graphene-SiO$_2$ contact is higher than that reported in literature \cite{Koenig2011,Jiang2015,Akinwande2016,Gao2017} or expected by theory \cite{Paek2013,Liu2014,Kumar2016}. Note that here we consider the bottom of the silicon cavity to be covered with a native oxide layer. This higher-than-expected adhesion energy can have its origin in graphene slippage or sliding \cite{PerezGarza2014}. If slipping takes place, the strain in the membrane would relax compared to that calculated based on ideal clamping, thus skewing the obtained value of $\Gamma$.
More indication to this latter effect can be seen from comparing Raman images of a suspended and a stuck drum obtained by fitting the G and 2D peaks (see Electronic Supplementary Information).

In conclusion, this work presents a technique for fast yield characterization that can be automatized for large scale testing of graphene mechanical devices.
We use a model that allows the selection of appropriate color filters to identify freestanding graphene devices under an optical microscope. We use the colorimetry technique to analyze more than 21000 drums that have underwent many thousands of pumping-venting cycles, and we show that a single scaling parameter ($D^4/g^3$), derived from structural mechanics, governs the survival probability of these 2D suspended structures. Moreover, doubling the number of graphene layers increases significantly the device yield for a same value of this scaling parameter.
Furthermore, by analyzing the properties of DLG structures that are stuck to the bottom of the cavities, we are able to extract a value of adhesion energy as $\Gamma =$ 0.9~J/m$^2$. This value is larger than what has been reported in literature, and it could be skewed by graphene slippage.

This work provides a framework for studying yield and failure mechanisms in arrays of suspended 2D material membranes. In particular these methods are expected to be useful for yield optimization in applications that require VLSI 2D nanomechanical arrays such as sensing systems or graphene interferometric modulation displays (GIMOD).

\section*{Acknowledgments}
The authors thank Joseph Scott Bunch, Menno Poot and Robin Joey Dolleman for discussions. The research leading to these results has received funding from the European Union's Horizon 2020 research and innovation programme under grant agreement No 649953 (Graphene Flagship).

\bibliography{Bibliography.bib}
\bibliographystyle{plain}

\newpage

\setcounter{figure}{0}
\setcounter{table}{0}
\renewcommand\thefigure{S-\arabic{figure}}
\renewcommand\thetable{S-\arabic{table}}

\section*{Electronic Supplementary Information}

Electronic Supplementary Information Outline:

1. Fabrication of SLG and DLG drumheads.

2. Derivation of the expression for adhesion energy.

3. Raman spectroscopy of suspended and collapsed DLG drumheads.

4. Theoretical limit of large-diameter drums before collapse.

5. Stuck drums and the diameter-dependence of their occurrence probability.

6. Complete dataset of yield for SLG and DLG samples.

\section*{1. Fabrication of SLG and DLG drumheads.}

Single-layer graphene was grown by  chemical vapour deposition (CVD) using a 4'' cold wall reactor (Aixtron BM). Copper foil was used as the catalyst and a surface pre-treatment was carried out in order to remove the native copper oxide and other impurities. The synthesis was carried out at $1000^{\circ}$C using methane as the carbon source. After the synthesis, the single-layer graphene (SLG) was coated with a polymer layer and stacked onto a second single-layer graphene by using a semi-dry transfer process. The stacked CVD double-layer graphene (DLG) was transferred onto 5$\times$5~mm$^2$ SiO$_2$/Si substrates containing circular cavities by following a semi-dry transfer procedure. Finally, the supporting polymer layer was removed by annealing at $450^{\circ}$C for 2~hours in N$_2$ atmosphere.

\section*{2. Derivation of the expression for adhesion energy.}

We approximate the stuck region of the drum to a circular region concentric with the circular drum, and assume perfect clamping as shown in Fig. S1.
\begin{figure}[h]
	\centering
	\includegraphics[]{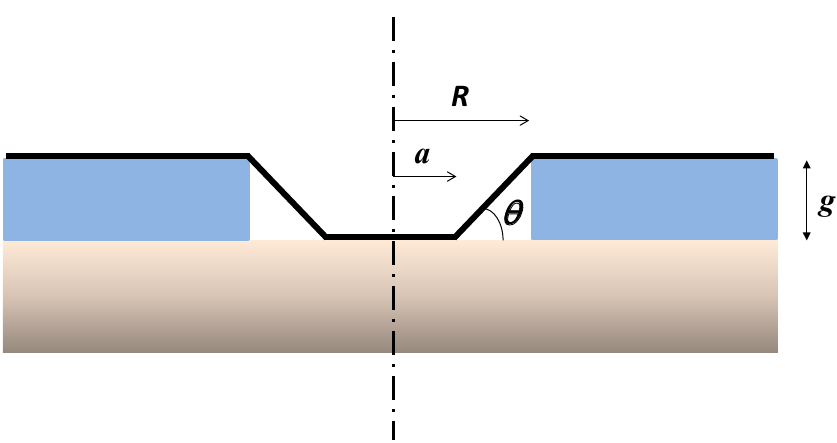}
	\caption{Model of a stuck circular membrane of radius $R$, the stuck area is assumed to be circular with a radius $a$, and a cavity depth $g$. The tensed suspended part of the membrane forms an angle $\theta$.}
	\label{fgr:figS1}
\end{figure}

The geometric proportions mean the angle $\theta$ is relatively small, we can therefore approximate the slope by:
\begin{equation}
\theta = \sin (\theta) \approx \tan(\theta) = \frac{g}{R-a} = \left(\frac{g}{R}\right) \frac{1}{1-\sqrt{\eta}}.
\label{eqn:EqS1}
\end{equation}
where $\eta$ is ratio of stuck-to-total area, $g$ is the cavity depth, $R$ is the drum's radius. We neglect radial displacement and assume only a vertical displacement $W(r)$ given as:
\begin{equation}
W(r) =
\begin{cases}
-g,    &\mathrm{for }~ 0\leq{r}<a \\
\frac{g}{R-a}r-g, &\mathrm{for }~ a\leq{r}\leq{R}
\end{cases}
\end{equation}

The radial strain induced by elongation can be obtained as [1]:
\begin{equation}
\epsilon_r =  \frac{1}{2} \left(\frac{\partial W(r)}{\partial r}\right)^2,
\label{eqn:EqS3}
\end{equation}
\noindent
The strain energy of the membrane is given by [1]:
\begin{equation}
U_s =  \frac{\pi E_{2D}}{2} \int_{0}^{R}\epsilon_r^2 rdr = \frac{\pi E_{2D}}{2} \int_{a}^{R}\epsilon_r^2 rdr,
\label{eqn:EqS4}
\end{equation}
Since the adhesion area does not change upon pressure cycling, the stuck drum is in a stable force equilibrium between adhesion force and the tension. The adhesion force given as [2]:
\begin{equation}
F_{ad} = 2 \pi a \Gamma,
\label{eqn:EqS4}
\end{equation}
\noindent
where $\Gamma$ is the adhesion energy in J/m$^2$.
Inserting equation 2 in equations 3 and 4, and applying the static force equilibrium condition, i.e. $F_{strain} = \frac{dU_s}{da} = F_{ad}$, the adhesion energy can be expressed in terms of geometric parameters and $E_{2D}$ as:
\begin{equation}
\Gamma = 16E_{2D} \left( \frac{g}{D}\right)^4 \frac{2+\sqrt{\eta}}{\sqrt{\eta} \left(1-\sqrt{\eta}\right)^4}.
\label{eqn:EqS2}
\end{equation}

\section*{3. Raman spectroscopy of suspended and collapsed drumheads.}
A Renishaw in via system is used to scan a focused laser spot ($\lambda$ = 514 nm) over the a suspended and a collapsed drumheads of 15~$\mu$m in diameter and 570~nm of cavity depth. We use the streaming feature to perform line scans of about 25~$\mu$m in 360~s with steps of 600~nm at a laser power of 25~mW.

\begin{figure}[h]
	\centering
	\includegraphics[]{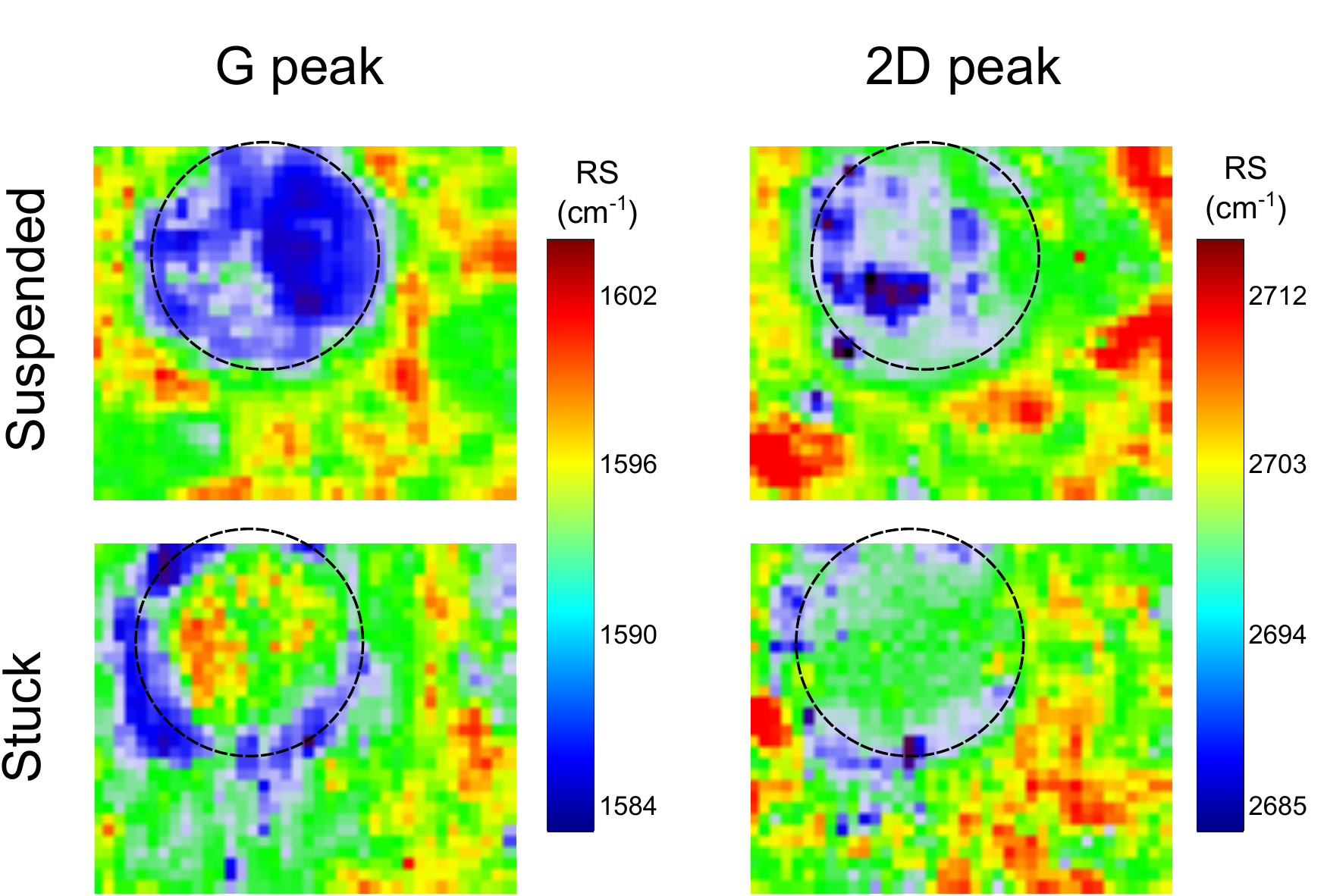}
	\caption{G and 2D Raman peaks maps of a freestanding (top) and a collapsed drums (bottom) of 13~$\mu$m in diameter.}
	\label{fgr:figS2}
\end{figure}

Figure S2 show the spatial maps of the G and 2D peaks of a freestanding (top) and a collapsed drums (bottom). These images not only indicate a high strain in the stuck drum (in the suspended areas), but also a higher strain in the anchoring area nearest to the drum compared to the same area of the suspended drum, thus indicating that some of the stress induced by stiction is being transferred to the graphene outside the drum.

\section*{4. Theoretical limit of large-diameter drums before collapse.}

Equation~3 in the main text relates the deflection $\delta$ of a circular membrane under a pressure difference $\Delta P$ across it:
\begin{equation}
\Delta P = \frac{16 T_0}{D^2} \delta + \frac{128 \pi E_{2D}}{3 D^4 (1-\nu) A_T} \delta ^3,
\label{eqn:EqS1}
\end{equation}

\noindent
where $T_0$ is the pretension in the structure, $E_{2D}$ is the 2D Young's modulus of the graphene membrane, $\nu = 0.16$ is the Poisson's ratio of graphene, and $D$ is the membrane's diameter.

Assuming a perfectly impermeable membrane with $T_0=0.017$~N/m (Cartamil-Bueno et al, work in progress) under a pressure difference of 1~bar, we can find the minimum value of the Young's modulus to deflect the center of the circular membrane a distance equal to the depth of its cavity $\delta = g$. Figure S3 shows the colormap of the minimum Young's modulus for different cavity depths and diameters in the case of SLG (left) and DLG (right).

\begin{figure}[h]
	\centering
	\includegraphics[]{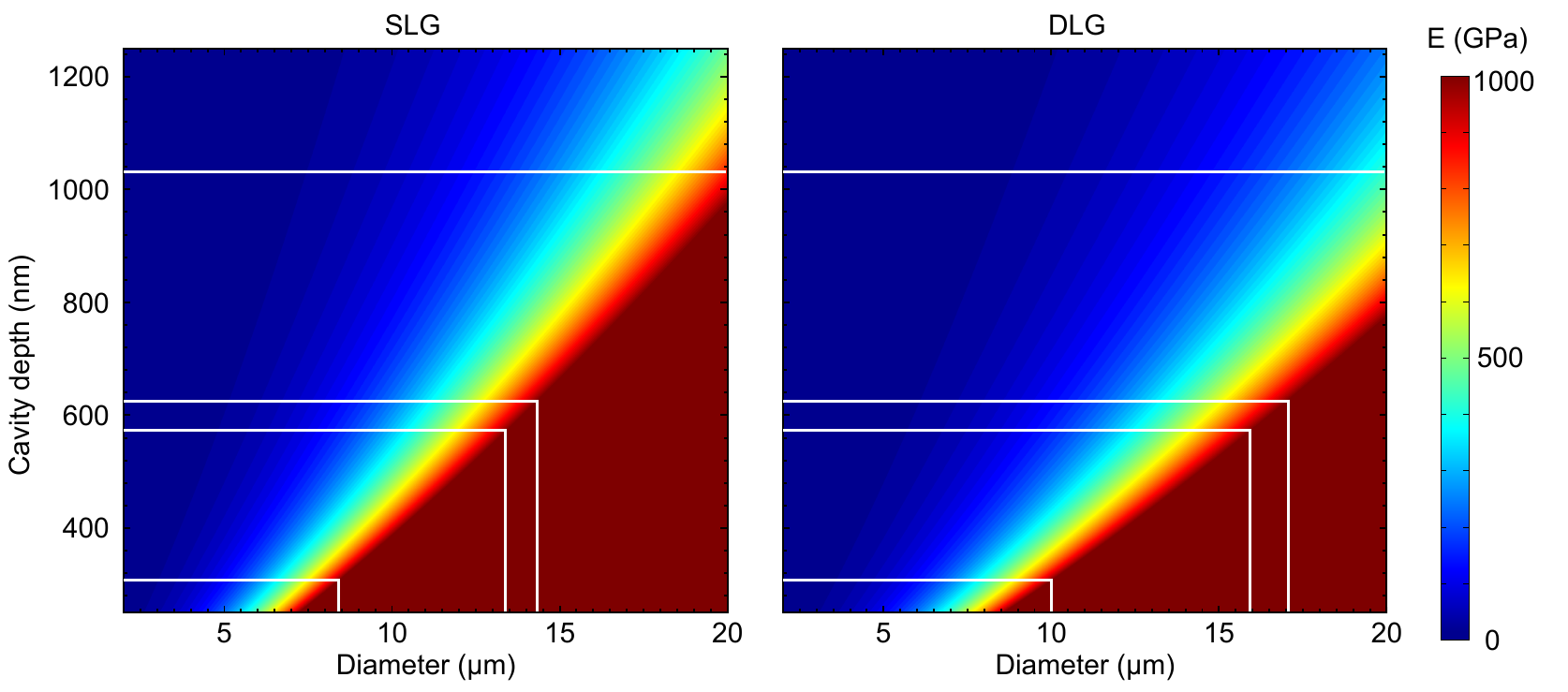}
	\caption{Minimum Young's modulus of circular SLG (left) and DLG (right) membranes for different diameters and cavity depths for which a pressure difference of 1~bar would induce a deflection equal to the cavity depth.}
	\label{fgr:figS3}
\end{figure}

Given the fact that pristine graphene has a Young's modulus of 1~TPa and assuming that is the maximum value a CVD graphene can have, the theoretical model predicts a maximum diameter for a particular cavity depth beyond which the membrane would contact the bottom of the cavity and possibly collapse. For instance, all DLG membranes larger than 10~$\mu$m in diameter would contact the bottom of their cavities for cavity depths of 285~nm. The fact that some of the drums survive (Figure~3 in the main text) even though they are expected to touch the cavity bottom according to Figure S-3, might be caused by unsticking from the cavity bottom or by membrane pores that result in smaller values of $\Delta P$.

\section*{5. Stuck drums and the diameter-dependence of their existence.}

Figure S-4 shows an AFM image of the drumhead membrane shown in the inset of Figure 5 in the main text.
Figure S-5 shows the ratio of stuck devices to the total number of devices as a function their diameter. Only samples with the four cavity depths indicated in the legend contain these stuck devices.

\begin{figure}[h]
	\centering
	\includegraphics[]{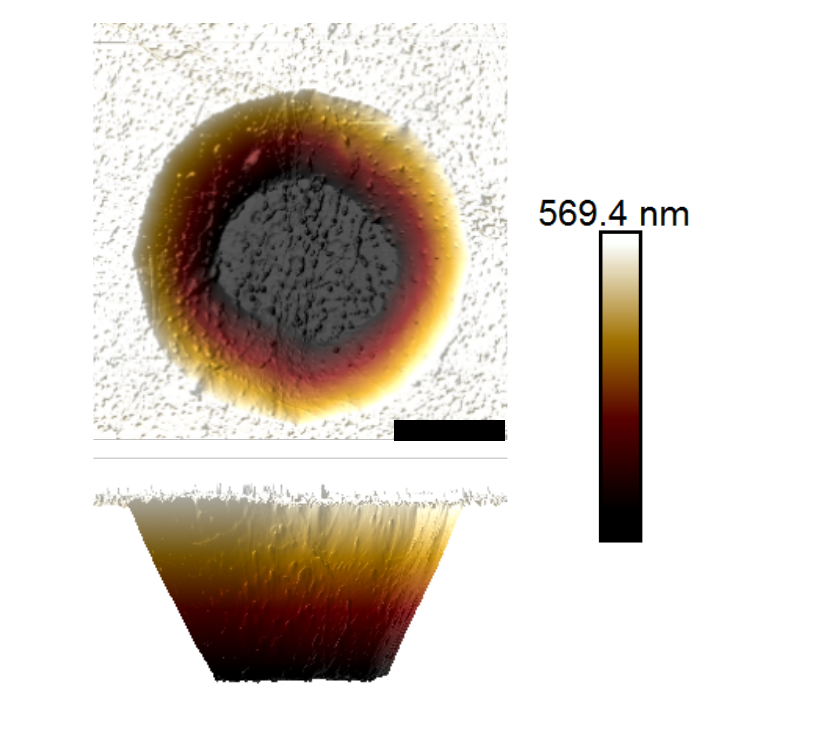}
	\caption{AFM image of a stuck drum see from above (top) and cross section (bottom). Scale bar is 5~$\mu$m.}
	\label{fgr:figS3}
\end{figure}

\begin{figure*}[h]
	\centering
	\includegraphics[]{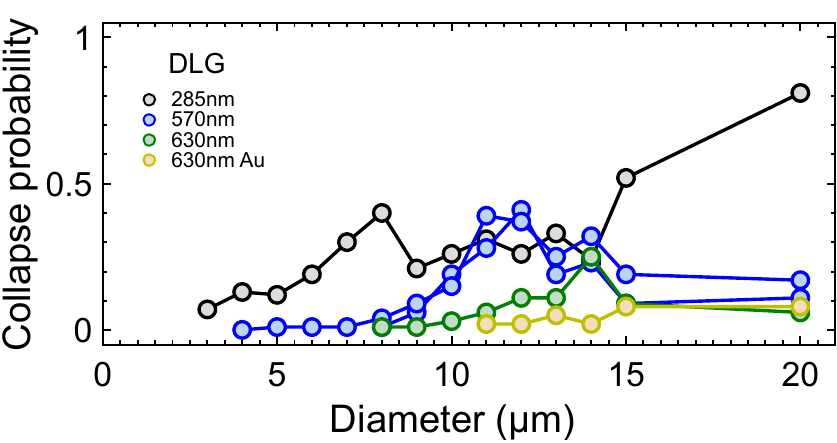}
	\caption{Proportion of stuck drums as a function of their diameter.}
	\label{fgr:figS3}
\end{figure*}

\section*{6. Complete dataset of yield for SLG and DLG samples.}
The tables in .xls files collect all the yield data for the SLG (table~S-1) and DLG samples (table~S-2) used in this work.

\section*{References}

[S-1] Yen, David HY, and T. W. Lee. "On the non-linear vibrations of a circular membrane." International Journal of Non-Linear Mechanics 10.1 (1975): 47-62.

\noindent
[S-2] Saif, M. T. A., B. Erdem Alaca, and Huseyin Sehitoglu. "Analytical modeling of electrostatic membrane actuator for micro pumps." Journal of Microelectromechanical Systems 8.3 (1999): 335-345.

\clearpage

\end{document}